# On the residues vectors of a rational class of complex functions. Application to autoregressive processes.

Guillermo Daniel Scheidereiter[i]    -    Omar Roberto Faure[ii]

July 10, 2019

## Abstract

Complex functions have multiple uses in various fields of study, so analyze their characteristics it is of extensive interest to other sciences. This work begins with a particular class of rational functions of a complex variable; over this is deduced two elementals properties concerning the residues and is proposed one results which establishes one lower bound for the p-norm of the residues vector. Applications to the autoregressive processes are presented and the exemplifications are indicated in historical data of electric generation and econometric series.

## 1. Choice of Theme

Consider the class of complex rational functions:

$$f: U \subset \mathbb{C} \to \mathbb{C} \text{ such that } f(z) = \frac{1}{1 - \alpha_1 z - \alpha_2 z^2 - \cdots - \alpha_n z^n} \quad (1)$$

that meet the following conditions:

(I) $\alpha_n \neq 0$ and $|\alpha_1| + |\alpha_2| + \cdots + |\alpha_n| < 1$ with $\alpha_j \in \mathbb{R} \ for \ j = 1, \ldots, n;$

(II) The poles $z_1, z_2, \ldots, z_n$ of $f(z)$ are simple poles that satisfy $|z_j| > 1$ for $j = 1, \ldots, n.$

Be the subclass of functions of (1):

$$f(z) = \frac{1}{1 - \alpha_1 z} \quad (2)$$

Here, $|\alpha_1| < 1$ and $\alpha_1 \neq 0$. The pole of $f(z)$ is $z_1 = \frac{1}{\alpha_1}$ and in such a case, the residue at the pole is calculated by:

$$\operatorname*{Res}_{z_1=\frac{1}{\alpha_1}} f(z) = \lim_{z \to \frac{1}{\alpha_1}} \left( z - \frac{1}{\alpha_1} \right) f(z)$$

Result:

$$\operatorname*{Res}_{z_1=\frac{1}{\alpha_1}} f(z) = c_{-1}^1 = -\frac{1}{\alpha_1}.$$

Since $|\alpha_1| < 1$, is obtained:

$$|c_{-1}^1| > 1 \quad (3)$$

By virtue of this immediate result over the subclass of complex functions whose denominator has degree $n = 1$, it is of interest to study the properties that can be obtained from the analysis of the behavior of the residues of more general cases than (2).

## 2. Definition of the problem

The object of study is the residuals $c_{-1}^j$ for $j = 1, \dots, n$ of the type of complex functions of the form defined in (1) that satisfy the conditions (I) and (II). These functions will have as many poles as indicates the degree the polynomial of the denominator and, therefore, a series of Laurent associated to each pole. Starting from (3), is analyzed the pattern followed by the residuals of these complex functions for more general cases than (2) and it is proposed to answer what properties they fulfill in relation to vectorial norms.

## 3. Conceptual theoretical framework

An open set[iii] $\Delta$ in the complex plane is *connected* if in any division of the same in two non-empty subsets, without common elements, $\Delta_1$ and $\Delta_2$, at least one of these sets contains a point of accumulation of the other set (Markushevich, 1970). Also, if a set is open and connected, it is a domain (Churchill & Brown, 1992).

In the geometry of the complex plane, the equation of an arc $\gamma$ is given by $z(t) = x(t) + iy(t)$ where $a \leq t \leq b$ and $x(t), y(t)$ are continuous real functions. A bow is simple, or Jordan arch, if $z(t_1) = z(t_2)$ only for $t_1 = t_2$. Also, an arc is a closed curve if its end points coincide $z(a) = z(b)$ (Ahlfors, 2013). A curve $\gamma$ is regular to pieces in $[a, b]$ if it has a bounded derivative in all $[a, b]$ except (perhaps) in a finite number of points (Apostol, 2009).

Let $f(z)$ be an *analytic*[iv] function in an annular domain $r < |z - z_0| < R$, and let $\gamma$ be any simple closed contour around $z_0$, positively oriented, contained in that domain. Then, at every point $z$ of that domain, $f(z)$ admits the serial representation:

$$f(z) = \sum_{n=-\infty}^{\infty} c_n(z - z_0)^n \quad (4)$$

where

$$c_n = \frac{1}{2\pi i} \int_\gamma \frac{f(z)dz}{(z - z_0)^{n+1}} \quad (n = 0, \pm 1, \pm 2, \dots) \quad (5)$$

The form (4) is called the *Laurent series* (Churchill & Brown, 1992).

On the other hand, if $f$ is an *analytic* function in $|z - z_0| < R$, it is said that $f$ has in $z_0$ a *zero* of order $m$ if $f(z) = (z - z_0)^m g(z)$, where $g$ is *analytic* in $z_0$ and $g(z_0) \neq 0$.

Consider now a function that is analytic in an neighborhood of $z_0$, except possibly in the same $z_0$. This is, $f(z)$ is analytical in the region $0 < |z - z_0| < R$. The point $z_0$ is called an isolated singularity of $f(z)$. If $\lim_{z \to z_0} f(z) = \infty$, it is said that the point $z_0$ is a *pole*[v] of $f(z)$ (Ahlfors, 2013).

The coefficient $c_{-1}$ that multiplies $(z - z_0)^{-1}$ in the development given in (4) for $f(z)$, that is obtained from (5) making $n = -1$, it's called *residue* of $f$ in $z_0$:

$$c_{-1} = \operatorname*{Res}_{z = z_0} f(z) = \frac{1}{2\pi i} \int_\gamma f(z) \, dz. \quad (6)$$

In practice, (6) is calculated according to the order of the poles of the function. In particular, if $f$ has a simple pole[vi] in $z_0$, then

$$\operatorname*{Res}_{z = z_0} f(z) = \lim_{z \to z_0} (z - z_0) f(z). \quad (7)$$

If $\gamma$ is a closed curve regular to pieces whose graph does not contain $z_0$, then

$$n(\gamma, z_0) = \frac{1}{2\pi i} \int_\gamma \frac{dz}{z - z_0} \quad (8)$$

it is called the number of turns (or index) of $\gamma$ with respect to $z_0$. Suppose, on the other hand, that $f(z)$ is *analytic* in an open disk $\Delta^{\text{vii}}$, and that $\gamma$ is a closed curve in that open. For some point $z_0$ that does not belong to $\gamma$

$$n(\gamma, z_0) f(z_0) = \frac{1}{2\pi i} \int_\gamma \frac{f(z)}{z - z_0} dz \quad (9)$$

where $n(\gamma, z_0)$ is defined by (7) (Ahlfors, 2013). When $n(\gamma, z_0) = 1$, the most frequent form of use is obtained:

$$f(z_0) = \frac{1}{2\pi i} \int_\gamma \frac{f(z)}{z - z_0} dz. \quad (10)$$

In Complex Analysis, (9) and (10), they are known as the Cauchy integral formula.

On the other hand, the set $\mathbb{C}^n = \mathbb{R}^n + i\mathbb{R}^n$ is the $n$-dimensional complex vector space that consists of all vectors $\boldsymbol{z} = \boldsymbol{x} + i\boldsymbol{y}$, where $\boldsymbol{x}, \boldsymbol{y} \in \mathbb{R}^n$ and $i$ is the usual imaginary unit that satisfies $i^2 = -1$ (Scheidemann, 2005). The points or vectors of $\mathbb{C}^n$ can be written in the form $\boldsymbol{z} = (z_1, \ldots, z_n) = \boldsymbol{x} + i\boldsymbol{y} = (x_1 + iy_1, \ldots, x_n + iy_n)^{\text{viii}}$. Let $p \in \mathbb{N}$ be a natural number $\geq 1$. The following expressions define norms[ix] in $\mathbb{C}^n$:

$$\|\boldsymbol{z}\|_\infty = \max_{1 \leq j \leq n} |z_j| \quad (11)$$

$$\|\boldsymbol{z}\|_p = \left( \sum_{j=1}^n |z_j|^p \right)^{\frac{1}{p}} \quad (12)$$

$\|\boldsymbol{z}\|_\infty$ it is called maximum norm (or infinity) and $\|\boldsymbol{z}\|_p$ is the $p-$norm (Scheidemann, 2005). It can be proven that $\|\boldsymbol{z}\|_\infty = \lim_{p \to \infty} \|\boldsymbol{z}\|_p$ that for every vector $\boldsymbol{z}$ in $\mathbb{C}^n$. Also,

$$\|\boldsymbol{z}\|_\infty \leq \|\boldsymbol{z}\|_2 \leq \sqrt{n} \|\boldsymbol{z}\|_\infty. \quad (13)$$

## 4. Justification of the study

Taking into account function (1) with the conditions (I) and (II), the subclass of functions where the denominator is of degree $n = 2$ satisfy the following result:

If

$$f(z) = \frac{1}{1 - \alpha_1 z - \alpha_2 z^2}, \quad (14)$$

then

$$\left| c_{-1}^j \right| > \frac{1}{2} \text{ for } j = 1,2. \quad (15)$$

In effect, the function has two poles that depend on the coefficients $\alpha_1$ and $\alpha_2$:

$$z_{1,2} = \frac{\alpha_1 \pm \sqrt{\alpha_1{}^2 + 4\alpha_2}}{-2\alpha_2}.$$

In this case,

$$\operatorname*{Res}_{z=z_{1,2}} f(z) = c_{-1}^{1,2} = \lim_{z \to z_{1,2}} \left( z - \frac{\alpha_1 \pm \sqrt{\alpha_1{}^2 + 4\alpha_2}}{-2\alpha_2} \right) . f(z).$$

Developing algebraically:

$$c_{-1}^{1,2} = \pm \frac{1}{\sqrt{\alpha_1{}^2 + 4\alpha_2}}. \quad (16)$$

As $|\alpha_1| + |\alpha_2| < 1$, implies $|\alpha_1| < 1 - |\alpha_2|$:

$$\left| c_{-1}^{1,2} \right| > \frac{1}{\sqrt{(1 - |\alpha_2|)^2 + 4\alpha_2}} = \frac{1}{\sqrt{1 - 2|\alpha_2| + |\alpha_2|^2 + 4\alpha_2}}$$

Suppose that $0 < \alpha_2 < 1$. Then,

$$\left| c_{-1}^{1,2} \right| > \frac{1}{\sqrt{\alpha_2{}^2 + 2\alpha_2 + 1}} = \frac{1}{\sqrt{(\alpha_2 + 1)^2}}$$

Consequently,

$$\left| c_{-1}^{1,2} \right| > \frac{1}{|\alpha_2 + 1|}.$$

Since $0 < \alpha_2 < 1$, it is clear that $|\alpha_2 + 1| < 2$, so that $\frac{1}{|\alpha_2+1|} > 0.5$. Then,

$$\left|c_{-1}^{1,2}\right| > \frac{1}{2}.$$

Similarly, it can be studied for $-1 < \alpha_2 < 0$.

From (15):

$$\left(c_{-1}^{j}\right)^2 > \frac{1}{4}.$$

Adding each residue squared:

$$(c_{-1}^1)^2 + (c_{-1}^2)^2 > \frac{1}{2}$$

Taking square root on both sides of the inequality:

$$\sqrt{(c_{-1}^1)^2 + (c_{-1}^2)^2} > \frac{1}{\sqrt{2}}$$

Since $\sqrt{(c_{-1}^1)^2 + (c_{-1}^2)^2} = \|\boldsymbol{c_{-1}}\|_2$, where $\boldsymbol{c_{-1}} = (c_{-1}^1, c_{-1}^2)$, that is, the vector whose components are the residuals of function (14) at the poles $z_1$ and $z_2$, is obtained:

$$\|\boldsymbol{c_{-1}}\|_2 > \frac{1}{\sqrt{2}}. \quad (17)$$

The above shows that the norm of the residues vector of function (14) has a lower bound. Furthermore, $|c_{-1}^1| + |c_{-1}^2| > \frac{1}{2} + \frac{1}{2}$ and from (12):

$$\|\boldsymbol{c_{-1}}\|_1 > 1. \quad (18)$$

On the other hand, if $|c_{-1}^j| > \frac{1}{2}$ for $j = 1,2$, then $\max_{j=1,2}|c_{-1}^j| > \frac{1}{2}$. That is,

$$\|\boldsymbol{c_{-1}}\|_\infty > \frac{1}{2}. \quad (19)$$

Therefore, the need to study which general properties meet the residuals of the class of functions (1) for a denominator with degree $n$ is justified. In particular, the $p$-norm of the residues vector is analyzed for the function defined in (1) with conditions (I) and (II).

## 5. Limitations

Suppose we consider the function:

$$f(z) = \frac{1}{1 - 2z + 3z^2},$$

which satisfies (14) as a particular form of (1). In such a case, the residue vector is:

$$\boldsymbol{c_{-1}} = \left( -\frac{\sqrt{2}}{4}i, \frac{\sqrt{2}}{4}i \right)_{1 \times 2}$$

Calculating the norms it is observed that they are not fulfilled (17), (18) and (19):

$$\|\boldsymbol{c_{-1}}\|_2 = \frac{1}{2} < \frac{1}{\sqrt{2}}$$

$$\|\boldsymbol{c_{-1}}\|_1 = \frac{\sqrt{2}}{2} < 1$$

$$\|\boldsymbol{c_{-1}}\|_\infty = \max_{j=1,2} |c_{-1}^j| = \frac{\sqrt{2}}{4} < \frac{1}{2}.$$

It is clear that $f(z)$ does not verify the condition (I), since

$$|2| + |-3| > 1.$$

Nor is condition (II) fulfilled. In effect, calculating the poles of the function:

$$z_1 = \frac{1}{3} + \frac{\sqrt{2}}{3}i \quad \text{and} \quad z_2 = \frac{1}{3} - \frac{\sqrt{2}}{3}i$$

it is immediate that $|z_1| < 1$ and $|z_2| < 1$.

The results studied here have implications for the functions of form (1) that strictly satisfy conditions (I) and (II). They do not extend to these functions when they do not verify (I) and/or (II) nor to other complex functions.

## 6. Scope of Work

Consider an autoregressive model of order $s$, $AR(s)$, with equation:

$$y_t = \alpha + \phi_1 y_{t-1} + \phi_2 y_{t-2} + \cdots + \phi_s y_{t-s} + A_t \quad (20)$$

where $A_t$ is a white noise process with zero mean and constant variance, $\{A_t\} \sim iid(0, \sigma^2)$, the parameters $\phi_1, \phi_2, \ldots, \phi_s$ are the coefficients of the model and $\alpha$ is a constant such that the mean $\mu$ of the model is:

$$\mu = \frac{\alpha}{1 - \phi_1 - \phi_2 - \cdots - \phi_s}. \quad (21)$$

The equation (20) can be written (Shumway & Stoffer, 2011):

$$y_t - \alpha = \phi_1(y_{t-1} - \alpha) + \phi_2(y_{t-2} - \alpha) + \cdots + \phi_s(y_{t-s} - \alpha) + A_t.$$

Doing $w_t = y_t - \alpha$ results:

$$w_t = \phi_1 w_{t-1} + \phi_2 w_{t-2} + \cdots + \phi_p w_{t-s} + A_t$$

Using the operator, $B^d w_t = w_{t-d}$ with $d \in \mathbb{Z}_{\geq 1}$:

$$(1 - \phi_1 B - \phi_2 B^2 - \cdots - \phi_s B^s)w_t = A_t \quad (22)$$

Here $\varphi_s(B) = 1 - \phi_1 B - \phi_2 B^2 - \cdots - \phi_s B^s$ is the autoregressive polynomial of order $s$, which can be written in terms of the complex variable $z$:

$$\varphi_s(z) = 1 - \phi_1 z - \phi_2 z^2 - \cdots - \phi_s z^s \quad (23)$$

From (22) the complex operator of the model is obtained:

$$\xi(z) = \frac{1}{1 - \phi_1 z - \phi_2 z^2 - \cdots - \phi_s z^s} \quad (24)$$

The process described by equation (20) is stationary when the roots of the equation $1 - \phi_1 z - \phi_2 z^2 - \cdots - \phi_s z^s = 0$ are outside the unit circle (Wei, 2006). For this to happen and the process to be stationary on average, of (21), it is sufficient that $|\phi_s| + \cdots + |\phi_1| < 1$. If in addition it is added as a condition that the roots of (23) are simple, then (24) satisfies the form (1) and fulfills the conditions (I) and (II).

Therefore, the vectorial properties of the residuals of the class of complex functions studied in this work, directly reach the autoregressive processes whose complex operators verify (I) and (II). For example, if you have an autoregressive process of order $s = 2$ stationary, with $|\phi_1| + |\phi_2| < 1$ and such that the roots of the equation $1 - \phi_1 z - \phi_2 z^2 = 0$ are simple, then the residues vector $c_{-1}^j$ for $j = 1,2$ of the associated complex operator

$$\xi(z) = \frac{1}{1 - \phi_1 z - \phi_2 z^2},$$

verifies the conditions (17), (18) and (19).

## 7. Objectives

As shown in (3) and (15), subclasses (2) and (14) of the general form (1), which satisfy (I) and (II), have residues with vector properties of interest. Specifically, it is shown that the remainder of subclass (2) is such that $|c_{-1}^1| > 1$ while the residues vector of subclass (14) verify (17), (18) and (19). We are looking, therefore, for a generalization that considers the norms of the residues vector of these complex functions for any degree of the denominator polynomial. In addition, it is important to show that these results are fulfilled in autoregressive models that are obtained as an explanation of the behavior of time series from real data.

## 8. Hypothesis

In the previous paragraphs it is shown that the complex functions considered here have associated residues vectors whose norms satisfy certain conditions for two particular cases. It is of interest, therefore, to formulate a conjecture that considers general orders for the residues vector norm. From the properties that meet the residuals of subclasses (2) and (14) of the complex functions of class (1) with conditions (I) and (II), the following result is proposed:

The vector $\boldsymbol{c_{-1}} = (c_{-1}^1, c_{-1}^2, \ldots, c_{-1}^n)$, where $c_{-1}^j$ for $j = 1, \ldots, n$ are the residuals at the poles $z_j$, of the function $f(z)$ defined in (1) with conditions (I) and (II), satisfies the following inequalities:

$$\|\boldsymbol{c_{-1}}\|_p > \left(\frac{1}{n}\right)^{\frac{p-1}{p}} \quad \text{and} \quad \|\boldsymbol{c_{-1}}\|_\infty > \frac{1}{n}.$$

## 9. Material and methods

As previously indicated, the study focuses on the properties that verify the vector norms whose elements are the residues of complex functions that satisfy the definition (1) with conditions (I) and (II). The hypothesis will be demonstrated using the inductive method in mathematics, according to which if a propositional function is true for $n = 1$ and of the truth for $h \in \mathbb{N}$ the truth for $h+1$ is deduced, then is concluded the truth for all $n \in \mathbb{N}$. On the other hand, for the applications the ARIMA[x] modeling methodology is used. Specifically, the data that make up the series of historical records of the Consumer Price Index and the price in Argentine Pesos of the Dollar are studied. These are taken from the official website of the National Institute of Statistics and Censuses (INDEC) of the Argentine Republic, www.indec.gob.ar. It is also exemplified with data of gross electricity generation (in MWh) that were taken from the official page of the Salto Grande Hydroelectric Dam, www.saltogrande.org and correspond to daily records.

## 10. Results

Let $f(z)$ be defined in (1) with conditions (I) and (II).

Suppose that $\Omega$ is a simply connected domain[xi] and that $\gamma$ is a simple closed contour in $\Omega$ that encloses the poles of $f(z)$ and the unit circle. If $f$ is *analytic* in $\Omega$ except for the zeros of $1 - \alpha_1 z - \alpha_2 z^2 - \cdots - \alpha_n z^n$, then

$$\|\boldsymbol{c_{-1}}\|_p > \left(\frac{1}{n}\right)^{\frac{p-1}{p}} \quad (25)$$

where $\boldsymbol{c_{-1}} = (c_{-1}^1, c_{-1}^2, \ldots, c_{-1}^n)$ and $c_{-1}^j$ for $j = 1, \ldots, n$ are the residues of $f(z)$ in the poles, $z_j$.

In effect, suppose that $\gamma$ is a simple closed contour enclosing within it the circle of radius one and to the circles $\gamma_1, \gamma_2, \ldots, \gamma_n$ which have center at the simple poles $z_1, z_2, \ldots, z_n$ of $f(z)$, respectively and such that each $\gamma_j$ for $j = 1, \ldots, n$, has a radius $R_j$ small enough to guarantee that the circles $\gamma_1, \gamma_2, \ldots, \gamma_p$ are mutually disjoint. The integral of $f(z)$ over each $\gamma_j$ is given by:

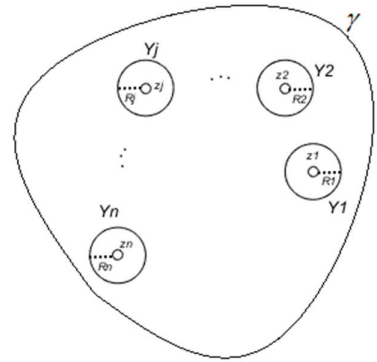

$$\int_{\gamma_j} f(z)\, dz = \int_{\gamma_j} \frac{dz}{1 - \alpha_1 z - \alpha_2 z^2 - \cdots - \alpha_n z^n} = \int_{\gamma_j} \frac{dz}{-\alpha_n(z - z_1) \ldots (z - z_j) \ldots (z - z_n)}$$

Rewriting the integrand, by the integral formula of Cauchy:

$$\int_{\gamma_j} f(z)\, dz = \int_{\gamma_j} \frac{\overline{\dfrac{1}{-\alpha_n(z-z_1)\ldots(z-z_n)}}}{z-z_j}\, dz = \int_{\gamma_j} \frac{g_j(z)}{z-z_j}\, dz = 2\pi i g_j(z_j)$$

where $g_j(z) = \frac{1}{-\alpha_n(z-z_1)\ldots(z-z_n)}$ and such that $z$ interior to $\gamma$, belongs to a region where there are no singularities of $g_j(z)$. On the other hand, let $\overline{p_l q_k}$ be a segment joining two points $p_l$ and $q_k$, with $l \neq k$, which are on $\gamma$ . Let $\delta = máx\{|\overline{p_l q_k}|\}$. Then[xii],

$$\left| g_j(z) \right| = \left| \frac{1}{-\alpha_n(z-z_1)\ldots(z-z_n)} \right| = \frac{1}{|\alpha_n|.|z-z_1|\ldots|z-z_n|} > \frac{1}{|\alpha_n|.\delta\ldots\delta} = \frac{1}{|\alpha_n|.\delta^{n-1}}.$$

Since[xiii] $|\alpha_n| < 1$, results:

$$\left| g_j(z) \right| > \frac{|\alpha_n|^n}{|\alpha_n|.\delta^{n-1}} = \left( \frac{|\alpha_n|}{\delta} \right)^{n-1}$$

The previous inequality is true for any $z$ in the domain of $g_j$, in particular for $z_j$. Therefore,

$$\left| g_j(z_j) \right| > \left( \frac{|\alpha_n|}{\delta} \right)^{n-1} \qquad (26)$$

Raising to an exponent $p \in \mathbb{N}$:

$$\left| g_j(z_j) \right|^p > \left( \frac{|\alpha_n|}{\delta} \right)^{p(n-1)}$$

Adding both sides for $j = 1, \ldots, n$:

$$\sum_{j=1}^{n} \left| g_j(z_j) \right|^p > n.\left( \frac{|\alpha_n|}{\delta} \right)^{p(n-1)}$$

Considering index root $p$:

$$\sqrt[p]{\sum_{j=1}^{n} \left| g_j(z_j) \right|^p} > \sqrt[p]{n}.\left( \frac{|\alpha_n|}{\delta} \right)^{n-1}$$

Clearly, the expression on the left side is the $p$-norm of the vector whose components are the values $g_j(z_j)$ for $j = 1, \ldots, n$. As

$$g_j(z_j) = \frac{1}{2\pi i} \int_{\gamma_j} f(z)\, dz = \operatorname*{Res}_{z=z_j} f(z) = c_{-1}^j$$

This is the $p$-norm of the residues vector:

$$\boldsymbol{c_{-1}} = (c_{-1}^1,\ c_{-1}^2, \ldots, c_{-1}^n)$$

Also, $\max_n\left\{\left(\frac{|\alpha_n|}{\delta}\right)^{n-1}\right\} = 1$ and it is reached for $n = 1$. In effect, the sequence $\{x_n\} = \left\{\left(\frac{|\alpha_n|}{\delta}\right)^{n-1}\right\}$ is bounded, since $|x_n| \leq \left|\frac{1}{\delta^{n-1}}\right|$ because $|\alpha_n| < 1$. In addition, $\delta \geq 2$, so that $|x_n| \leq \left|\frac{1}{2^{n-1}}\right|$. Since $\max_n\left\{\frac{1}{2^{n-1}}\right\} = 1$ and the terms of $\{x_n\}$ are bounded by the terms of $\left\{\frac{1}{2^{n-1}}\right\} = \left\{1, \frac{1}{2}, \frac{1}{4}, \frac{1}{8} \ldots\right\}$, is obtained $\max_n\left\{\left(\frac{|\alpha_n|}{\delta}\right)^{n-1}\right\} = 1$. Then, for $n = 1$:

$$\|\boldsymbol{c_{-1}}\|_p > 1. \quad (27)$$

What is (25) with $n = 1$. Next, the following proposition will be demonstrated:

$$\text{If } \|\boldsymbol{c_{-1}}\|_p > \left(\frac{1}{h}\right)^{\frac{p-1}{p}}, \text{ then } \|\boldsymbol{c_{-1}}\|_p > \left(\frac{1}{h+1}\right)^{\frac{p-1}{p}}$$

Suppose that $\|\boldsymbol{c_{-1}}\|_p \leq \left(\frac{1}{h+1}\right)^{\frac{p-1}{p}}$, where $\boldsymbol{c_{-1}} = (c_{-1}^1, c_{-1}^2, \ldots, c_{-1}^{h+1})$. Then,

$$\sqrt[p]{\sum_{j=1}^{h+1}\left|c_{-1}^j\right|^p} \leq \sqrt[p]{\left(\frac{1}{h+1}\right)^{p-1}}$$

And results:

$$\sum_{j=1}^{h+1}\left|c_{-1}^j\right|^p \leq (h+1)\cdot\frac{1}{(h+1)^p}.$$

The previous inequality can be written as follows:

$$\sum_{j=1}^{h}\left|c_{-1}^j\right|^p + \left|c_{-1}^{h+1}\right|^p \leq (h+1)\cdot\frac{1}{(h+1)^p}$$

It is immediate, therefore:

$$\sum_{j=1}^{h}\left|c_{-1}^j\right|^p \leq (h+1)\cdot\frac{1}{(h+1)^p}$$

Applying index root $p$ in both members:

$$\sqrt[p]{\sum_{j=1}^{h}\left|c_{-1}^{j}\right|^{p}} \leq \frac{\sqrt[p]{h+1}}{h+1} = \left(\frac{1}{h+1}\right)^{\frac{p-1}{p}} \leq \left(\frac{1}{h}\right)^{\frac{p-1}{p}}.$$

But this last result contradicts the hypothesis and arose from supposing the negation of the thesis. Then,

$$\|c_{-1}\|_p > \left(\frac{1}{h+1}\right)^{\frac{p-1}{p}}. \quad (28)$$

From (27), $\|c_{-1}\|_p > 1$. From (28), if $\|c_{-1}\|_p > \left(\frac{1}{h}\right)^{\frac{p-1}{p}}$, then $\|c_{-1}\|_p > \left(\frac{1}{h+1}\right)^{\frac{p-1}{p}}$. For the principle of induction, for all $n \in \mathbb{N}$:

$$\|c_{-1}\|_p > \left(\frac{1}{n}\right)^{\frac{p-1}{p}}.$$

They are particular cases:

$$\|c_{-1}\|_1 > 1 \text{ and } \|c_{-1}\| > \frac{1}{\sqrt{n}}. \quad (29)$$

Furthermore, $\lim_{p \to \infty}\left[\left(\frac{1}{n}\right)^{\frac{p-1}{p}}\right] = \lim_{p \to \infty}\left[\left(\frac{1}{n}\right)^{1-\frac{1}{p}}\right] = \frac{1}{n}$. Since $\|c_{-1}\|_\infty = \lim_{p \to \infty}\|c_{-1}\|_p$:

$$\|c_{-1}\|_\infty > \frac{1}{n}. \quad (30)$$

## 11. Application

The complex functions of form (1) appear in time series when they are modeled by autoregressive processes. A specific example is the series of daily electric generation records (MWh) of the Salto Grande Hydroelectric Dam, in the period from September 2018 to January 2019. The logarithm of the observed data and the estimation of the model were represented. From the information given by the correlogram of the simple and partial autocorrelation functions, the information criteria of Akaike, Schwarz and Hannan-Quinn and the tests based on the residuals[xiv] was obtained  the autoregressive model of order 1, $AR(1)$:

$$y_t = 4,17636 + 0,599419 y_{t-1} + A_t$$

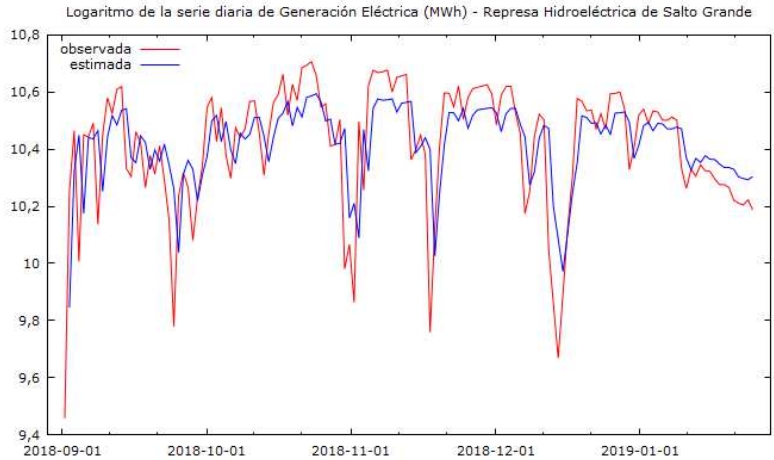

The operator (24) results:

$$\xi(z) = \frac{1}{1 - 0{,}599419z}.$$

This is subclass (2) of (1), where conditions (I) and (II) are satisfied. In such a case, $\alpha_1 = 0{,}599419 = \phi_1$:

$$c_{-1}^1 = -\frac{1}{\alpha_1} = -\frac{1}{0{,}599419} = -1{,}66828212.$$

Clearly, the autoregressive process is stationary and is fulfilled (3).

Another example is the time series corresponding to the Consumer Price Index, IPC, in Greater Buenos Aires. Specifically, the percentage variation of the index with respect to the previous month, in the period from January 1993 to December 2006, is taken. In this case, an autoregressive model of order $s = 2$ is estimated:

$$y_t = 0{,}584000y_{t-1} + 0{,}203494y_{t-2} + A_t$$

The following graph shows the observed variable and the one estimated by the model:

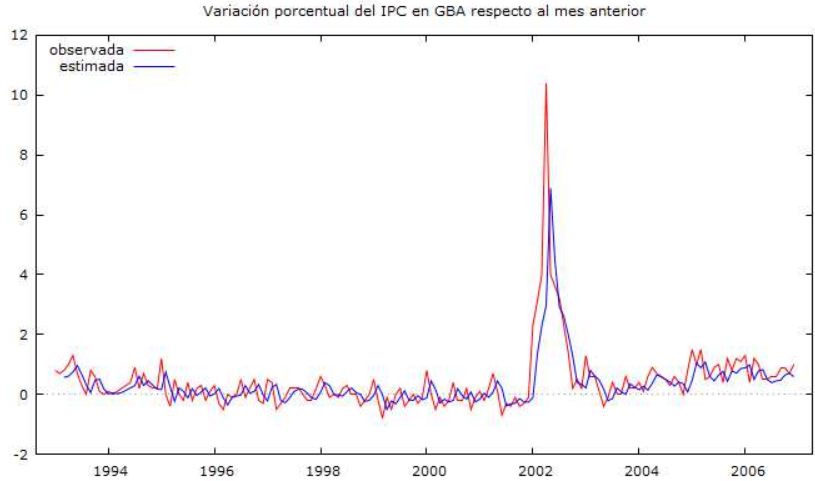

In this case, the associated complex operator is

$$\xi(z) = \frac{1}{1 - 0{,}584000z - 0{,}203494z^2}.$$

The function corresponds to the form (14) and conditions (I) and (II) are verified, because $|0{,}584000| + |0{,}203494| < 1$ and also, $|z_1| = |-4{,}0756094790| > 1$ and $|z_2| = |1{,}2057459941| > 1$. Using (16):

$$c_{-1}^{1,2} = \pm \frac{1}{\sqrt{(0{,}584000)^2 + 4.0{,}203494}} = \pm 0{,}9304713208.$$

The residues vector is:

$$\boldsymbol{c_{-1}} = (0{,}9304713208, -0{,}9304713208).$$

The norms verify (29) and (30):

$$\|\boldsymbol{c_{-1}}\|_2 = 1{,}31588516 > \frac{1}{\sqrt{2}}, \|\boldsymbol{c_{-1}}\|_1 = 1{,}86094264 > 1 \text{ and } \|\boldsymbol{c_{-1}}\|_\infty = 0{,}93047132 > \frac{1}{2}.$$

The autoregressive processes also appear in practice when, when studying a series of time, volatility in the variance[xv] is found, which is known as conditional heteroskedasticity dependent on time. The model to study this behavior

$$A_t^2 = \alpha_0 + \alpha_1 A_{t-1}^2 + \alpha_2 A_{t-2}^2 + \cdots + \alpha_s A_{t-s}^2 + \varepsilon_t \quad (31),$$

Is an $AR(s)$ with $\varepsilon_t \rightsquigarrow N(0, \sigma_\varepsilon^2)$ and it is called $ARCH$ (*Autoregressive Conditional Heteroscedasticity*). For example, if we consider the logarithm of the evolution of the cost of the Dollar in Argentine Pesos from the beginning of 2016 until

July 2018, we obtain an $AR(2)$ model with a differentiation order that is an $ARIMA(2,1,0)$ model:

$$(1-B)(1-0{,}267728B+0{,}143342B^2)y_t = A_t$$

The following graph shows the observed and estimated variable:

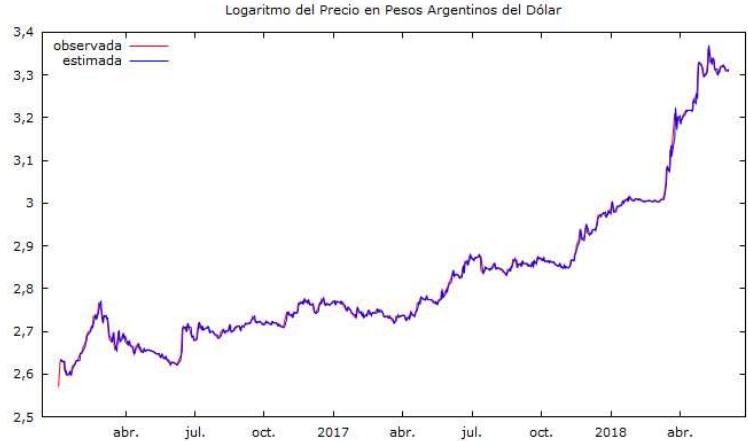

Evidence of conditional heteroskedasticity was found and modeled by an AR process (4). It turned out the complex operator:

$$\xi(z) = \frac{1}{1-0{,}128940z-0{,}116899z^2-0{,}153156z^3-0{,}169289z^4}$$

It is a complex function of class (1) with $n=4$ that verifies conditions (I) and (II). The residues vector is:

$$\boldsymbol{c_{-1}} = \begin{pmatrix} 0{,}291604006704114 - 0{,}300154215080589i \\ 0{,}291604006704114 + 0{,}300154215080589i \\ -0{,}291604006704114 - 0{,}241806751480026i \\ -0{,}291604006704114 + 0{,}241806751480026i \end{pmatrix}_{4\times 1}$$

From (25) the $p$-norm satisfies:

$$\|\boldsymbol{c_{-1}}\|_p > \left(\frac{1}{4}\right)^{\frac{p-1}{p}}.$$

In particular, $\|\boldsymbol{c_{-1}}\|_2 = 0{,}798284224250679 > \left(\frac{1}{4}\right)^{\frac{1}{2}} = 0{,}5$. Also, $\|\boldsymbol{c_{-1}}\|_\infty = 0{,}418479927304211 > 0{,}25$.

## 12. Conclusions

The first properties for the residuals of the complex functions (1) with the conditions (I) and (II), show for the subclass (2) that they are outside a circle of radius 1, while for (14) they remain outside a circle of radius $\frac{1}{2}$. In addition, the results (25) and (30), establish that the norms of the vector of residues of these complex rational functions have a lower bound that depends on the degree $n$ of the denominating polynomial and of the order $p$ of the norm that is considered:

$$\|c_{-1}\|_p > \left(\frac{1}{n}\right)^{\frac{p-1}{p}} \quad \text{and} \quad \|c_{-1}\|_\infty > \frac{1}{n}.$$

The $2$-norm and the infinite norm, when combined with (13), show another result to be considered:

$$\frac{1}{n} < \|c_{-1}\|_\infty \leq \|c_{-1}\|_2 \quad \text{and also} \quad \frac{\sqrt{n}}{n} < \|c_{-1}\|_2 \leq \sqrt{n}.\|c_{-1}\|_\infty.$$

That is, both standards have a higher level.

As shown in the discussion of the previous section, the complex functions (1) when they are fulfilled (I) and (II) have direct applicability over the autoregressive processes. In particular, they model the complex operator of a stationary process on average. Although in practice, the most frequent cases are $AR(1)$ and $AR(2)$ processes, the results (25) and (30) establish very general conditions on these processes whatever their order of autoregressivity, showing an alternative to study the stationarity of them. In other words, whenever a stationary autoregressive process has a complex operator that meets the conditions (I) and (II) of the rational class of complex functions of the form (1), will be met (25) and (30).

What is exposed in this paper points to an important connection between autoregressive processes and a certain class of complex rational functions. We will continue to investigate relationships that allow knowing alternative properties to those already known.

---

[i] Guillermo Daniel Scheidereiter, Facultad Regional Concordia, Universidad Tecnológica Nacional, Argentina, contact email: danielscheidereiter@gmail.com.

[ii] Omar Roberto Faure, Facultad Regional Concepción del Uruguay, Universidad Tecnológica Nacional, Argentina, contact email: ofaure@frcu.utn.edu.ar.

[iii] A set is *open* if it does not contain any of its border points.

[iv] The *class* of *analytic* functions consists of the complex functions of a complex variable that have a derivative wherever the function is defined. With the same meaning, the term *holomorphic* function is used (Ahlfors, 2013).

[v] If in (6), $c_{-n} \neq 0$ for some $n$ but $c_{-m} = 0$ for all $m > n$, the point $z_0$ it is called pole of order $n$. A pole of order 1 is usually called *simple pole* (Apostol, 2009).

[vi] For poles of order $n$ see, among others, (Zill & Shanahan, 2009).

[vii] Region of the complex plane defined by $|z - a| < R$.

[viii] Fort he vectors $\mathbf{z}$ and $\mathbf{w}$ in $\mathbb{C}^n$ the Euclidean norm and the internal product, respectively, are defined and used (Korevar & Wiegerinck, 2017): $\|\mathbf{z}\| = (|z_1|^2 + \cdots + |z_n|^2)^{\frac{1}{2}}$ and $\langle \mathbf{z}, \mathbf{w} \rangle = z_1 \bar{w}_1 + \cdots + z_n \bar{w}_n$. Clearly, the Euclidean norm is a particular case of (12) when $p = 2$.

[ix] $\mathbb{C}^n$ with the internal product (11) is a complex Hilbert space and the mapping $\mathbb{R}^n \times \mathbb{R}^n \to \mathbb{C}^n$, $(x, y) \to x + iy$ is an isometry. Thus, all metrics and topological notions in these spaces coincide. All rules define some topology in $\mathbb{C}^n$.

[x] The aforementioned bibliography can be consulted on this methodology.

[xi] A *simply connected* domain is a domain such that every simple closed contour within it encloses only points of it. When it is not simply connected it will be called *multiply connected* (Churchill & Brown, 1992).